\documentclass{emulateapj}

\slugcomment{}

\shorttitle{Microflare Activity driven by Forced Magnetic Reconnection}
\shortauthors{D.B. Jess et al.}

\begin{document}

\title{Microflare Activity driven by Forced Magnetic Reconnection}

\author{D. B. Jess, M. Mathioudakis}
\affil{Astrophysics Research Centre, School of Mathematics and Physics, Queen's University, Belfast, BT7~1NN, 
Northern Ireland, U.K.}
\email{d.jess@qub.ac.uk}

\author{P. K. Browning}
\affil{Jodrell Bank Centre for Astrophysics, School of Physics and Astronomy, University of Manchester, Manchester, 
M13~9PL, U.K.}


\author{P. J. Crockett and F. P. Keenan}
\affil{Astrophysics Research Centre, School of Mathematics and Physics, Queen's University, Belfast, BT7~1NN, 
Northern Ireland, U.K.}

\begin{abstract}
High cadence, multiwavelength, optical observations of a solar active region are presented, obtained with the 
Swedish Solar Telescope. Two magnetic bright points are seen to separate in opposite 
directions at a constant velocity of 2.8~kms$^{-1}$. After a separation distance of 
$\approx$4400~km is reached, multiple Ellerman bombs are observed in both H$\alpha$ and Ca-K images. 
As a result of the Ellerman bombs, periodic velocity perturbations in the vicinity of the 
magnetic neutral line, derived from simultaneous {\sc{mdi}} data, are generated with 
amplitude $\pm$6~kms$^{-1}$ and wavelength $\approx$1000~km. The velocity oscillations are followed 
by an impulsive brightening visible in H$\alpha$ and Ca-K, with a peak intensity enhancement of 63\%.  We interpret 
these velocity perturbations as the magnetic field deformation necessary to trigger forced reconnection.
A time delay of $\approx$3~min between the H$\alpha$-wing and Ca-K observations indicate that the observed 
magnetic reconnection occurs at a height of $\sim$200~km above the solar surface. These 
observations are consistent with theoretical predictions and provide the first observational evidence of 
microflare activity driven by forced magnetic reconnection.
\end{abstract}


\section{Introduction}
\label{intro}
Magnetic reconnection is a common phenomenon within the solar atmosphere. Its presence is often observed through explosive 
events such as solar flares, where extreme localized heating is generated through the conversion of magnetic energy 
(Priest~1986). Forced magnetic reconnection is a type of reconnection process which allows 
a fast transition of the magnetic configuration to a state of lower energy \citep[]{Vek98}.
In this regime, it is believed 
that the displacement of photospheric footpoints deform the initially smooth magnetic field configuration, generating a 
current sheet and forcing reconnection \citep[]{Jai05}. 
This form of reconnection is often modelled by implementing a 
sinusoidal boundary disturbance to a slab plasma with a field reversal, which drives the formation of a current sheet 
and causes reconnection at the magnetic neutral line \citep[]{Ma96}.

Observational evidence of magnetic reconnection occurring in the lower solar atmosphere (photosphere and chromosphere) 
is rare. It is generally assumed that reconnection and the subsequent release of magnetic energy is a coronal process, often 
occurring many thousands of km above the surface of the Sun. However, in recent years both \cite{Din99} and \cite{Che01} 
have provided evidence of eruptive phenomena and localized heating in the lower solar atmosphere, through the analysis of 
type~{\sc{ii}} white-light flares and Ellerman bombs. Unfortunately, for these cases there are currently no available 
theoretical models which can deal with the particle acceleration created. Indeed, since the surrounding plasma is much denser than in 
the corona, partially ionized and highly collisional, it is believed that most of the energy released through magnetic 
reconnection is promptly consumed \citep[]{Che06}. 
To date, there has been no observational evidence of flare activity driven by forced reconnection.

Theoretically, it was noted by \cite{Rou93} 
that reconnection could proceed rapidly in the photosphere due to the low 
conductivity of cooler plasmas, and this was used to interpret cancelling magnetic features as photospheric 
reconnections.  This idea was further developed by \cite{Lit99a} and \cite{Lit07}. 
However, the models presented are based on, respectively, oscillatory reconnection and steady-state Sweet Parker reconnection, 
which do not appear appropriate for transient events. 
In addition to the characteristic lower temperatures, reconnection in a partially-ionised 
photosphere may be distinguished from more commonly studied coronal reconnection by the interactions 
with neutrals. A few theoretical studies have tackled the question of reconnection in a partially-ionised 
plasma \citep[e.g.][]{Bul98}, 
but the process remains poorly understood. 

The data presented here are part of an observing sequence obtained on 2007 August 24, with the Swedish Solar Telescope (SST) 
on the island of La~Palma. Our optical setup allowed us to image a $68'' \times 68''$ region surrounding active 
region NOAA~10969 complete with solar rotation tracking. This active region was located at heliocentric 
co-ordinates ($-516''$,$-179''$), or S05E33 in the solar NS-EW co-ordinate system. The Solar Optical Universal 
Polarimeter (SOUP) was implemented to provide 2-dimensional spectral information across the H$\alpha$ line 
profile centred at $6562.8$~{\AA}. In addition, a series of Ca~{\sc{ii}} interference filters were used to provide 
high-cadence imaging in this portion of the optical spectrum. The observations and data reduction employed in the 
present analysis are described in detail by \cite{Jes08}. 
For the purpose of this letter, 2-dimensional H$\alpha$ 
spectral scans, Ca~K core and blue (Ca~{\sc{ii}}) continuum observations will be presented, along with full-disk 
{\sc{mdi}} magnetograms which are used to evaluate magnetic neutral line locations. During the 
observing sequence, a microflare event was observed, originating south-east of NOAA~10969 at 08:42~UT. 

Traditional microflares are defined as transient bursts, which release energy into the surrounding solar plasma 
over timescales normally less than 10~minutes \citep[]{Kru02}. 
These bursts, classified as low {\sl{GOES}} 
C class to below A class events, frequently produce hard X-rays as a result of Bremsstrahlung emission when 
non-thermal particles collide with the dense lower solar atmosphere \citep[]{Han08}. 
Additionally, microflares 
are well correlated with large-scale solar activity, and as a result only occur in the vicinity of active regions \citep[]{Chr08}. 
Therefore, while the term ``microflare'' is normally reserved for phenomena detectable in X-rays, on this occasion 
no such emission was established. 
However, due to consistencies with typical microflare characteristics (as will be discussed throughout), we deem this 
event an ``H$\alpha$ microflare''. In this letter we utilize the high-cadence multiwavelength data set 
to search for definitive evidence of microflare activity in the lower solar atmosphere which is driven 
by forced reconnection.

\section{Analysis and Results}
\label{analy}

\begin{figure*}
\epsscale{1.0}
\plotone{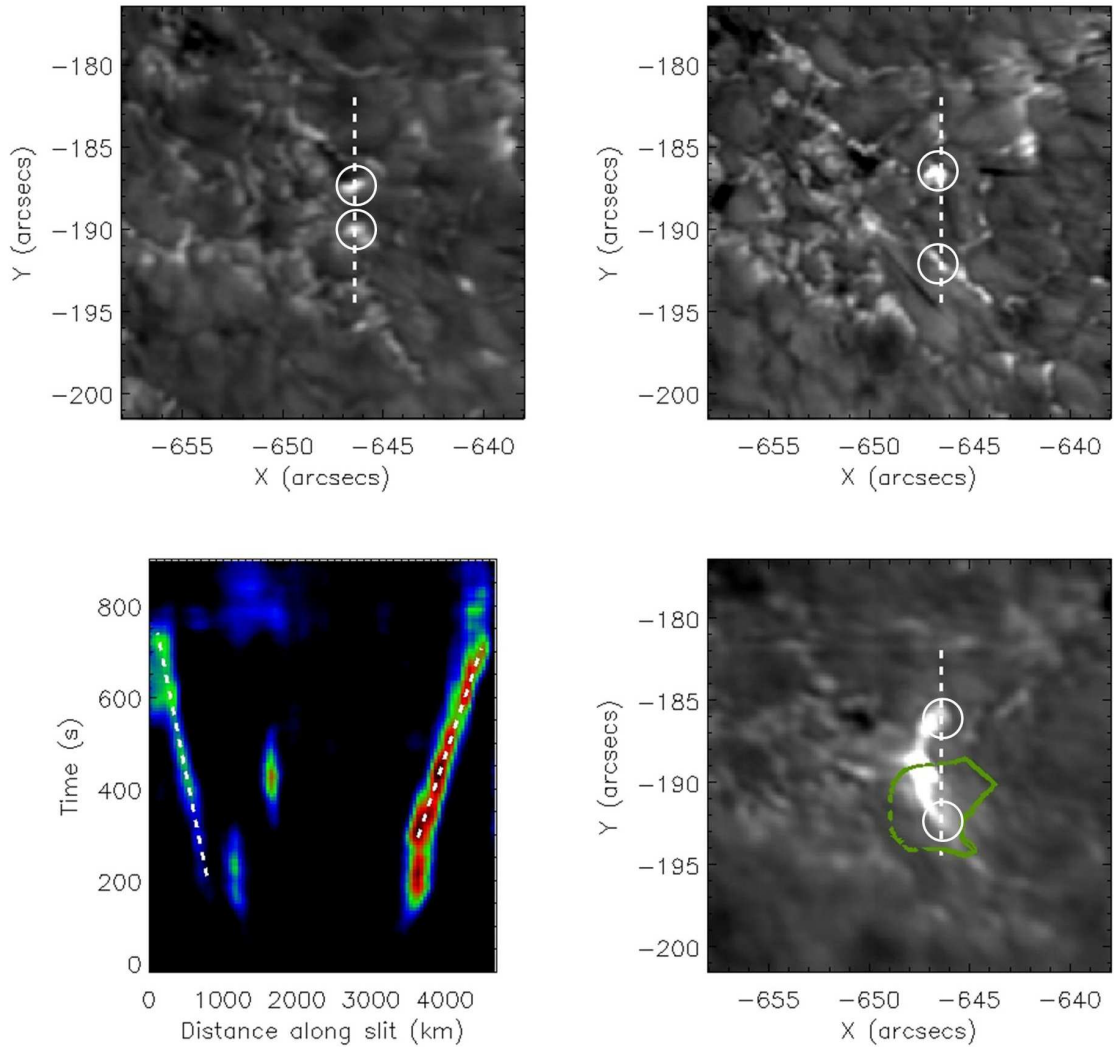}
\caption{{\it{Top Left:}} The separation of the MBPs at 08:28~UT. {\it{Top Right:}} The greater MBP separation 
observed at 08:38~UT. White circles mark the expected position of the footpoints and the dashed white 
lines indicate the slit position used to establish velocities. {\it{Lower Left:}} Time-distance plot 
demonstrating the observed separation of the MBPs as a function of time. The dashed 
white lines indicate a least-squares fit to the slopes, with the gradient providing the 
velocity for each MBP. The gradient observed indicates the presence of constant separation 
velocities, evaluated to be 1.0~kms$^{-1}$ and 1.8~kms$^{-1}$ for the left (most northerly) and right (most southerly) 
MBPs, respectively. {\it{Lower Right:}} H$\alpha$ microflare event, observed at the time of peak emissive flux (08:42~UT), 
with green contours marking the location of the magnetic neutral line. The brightest 
emission from the loop structure, seen to link the two MBPs, is co-spatial with the underlying neutral line. 
\label{f1}}
\end{figure*}

Prior to the H$\alpha$ microflare event, two magnetic bright points \citep[MBPs;][]{Cro09} 
are observed at 08:28~UT to be 
separated by $\approx$2700~km at heliocentric coordinates ($-646''$, $-188''$), as indicated by the white 
circles in the upper panels of Figure~\ref{f1}. Over the next 10~min, the MBPs are observed in both the blue 
Ca~{\sc{ii}} continuum and H$\alpha$-wing (core $\pm$ 700~m\AA) observations to diverge until 
they are separated by a distance exceeding $\approx$4400~km. The lower-left panel of 
Figure~\ref{f1} is a 
time-distance plot, obtained by placing a slit along the motion paths of the MBPs, 
revealing the displacement 
of these structures as a function of time. By establishing the gradient, and hence velocity, associated with 
the MBPs in the time-distance plot, as indicated by the dashed white 
lines in the lower-left panel of Figure~\ref{f1}, it is clear that each MBP displays a constant separation velocity, equal to 
$\approx$1.0~kms$^{-1}$ for the most northerly MBP and $\approx$1.8~kms$^{-1}$ for the most southerly one. Due to 
the movement of the two MBPs in opposite directions, the net separation velocity equates to 
$\approx$2.8~kms$^{-1}$. This velocity is consistent with the measured separation distance over the 10~min duration 
($\frac{4400~\mathrm{km}~-~2700~\mathrm{km}}{600~\mathrm{s}}$ $=$ 2.83~kms$^{-1}$), and reiterates the presence of a constant separation 
velocity in opposing directions.

\begin{figure*}
\epsscale{1.0}
\plotone{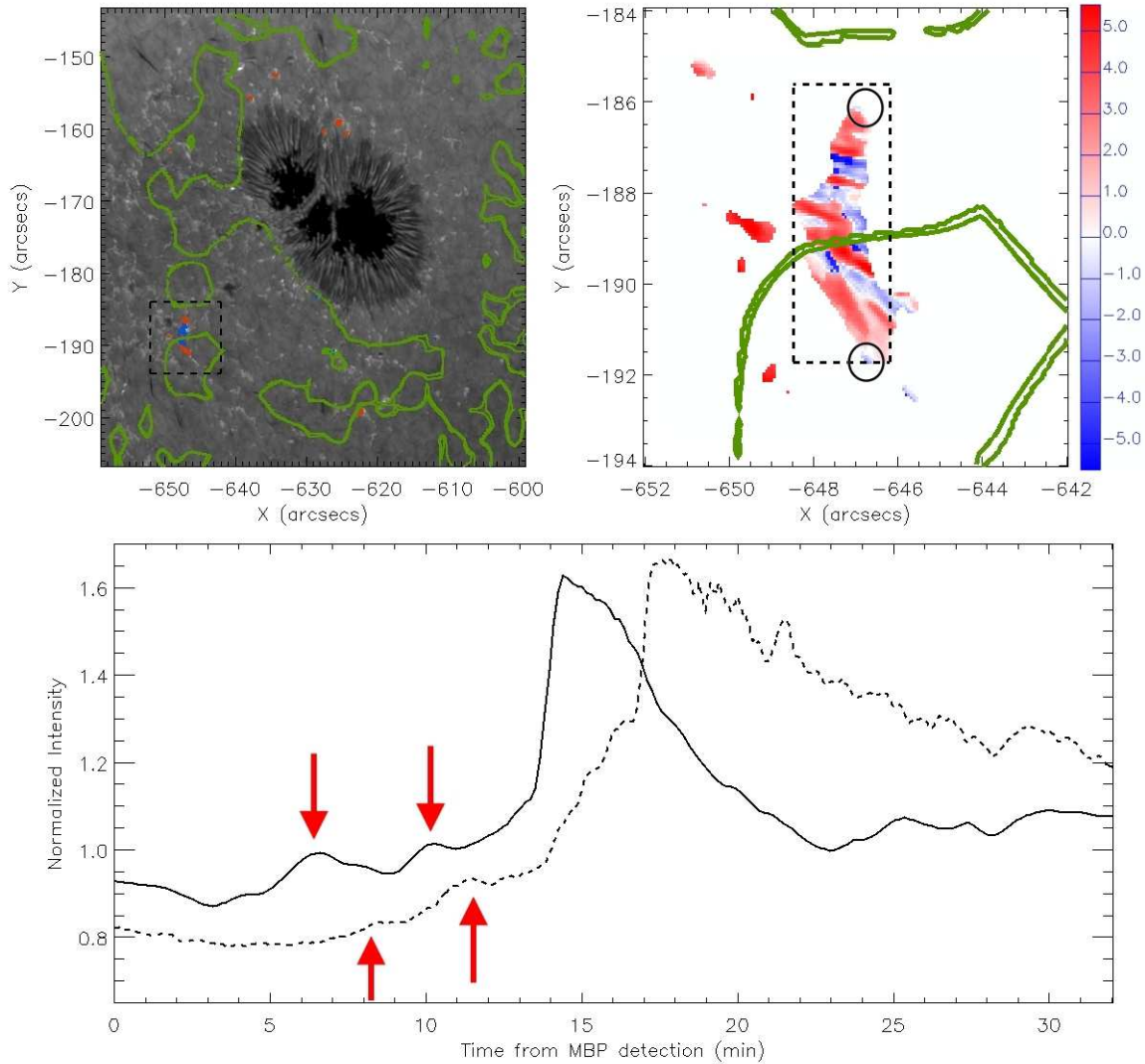}
\caption{{\it{Upper Left:}} H$\alpha$ blue-wing (H$\alpha$-core $-700$~m{\AA}) field-of-view overplotted with velocity contours 
exceeding $\pm$3~kms$^{-1}$. Red contours indicate red-shifted Doppler velocities, blue contours indicate 
blue-shifted Doppler velocities, and green contours mark the locations of the magnetic neutral line. 
The dashed black lines mark a region which is magnified in the upper-right panel. 
{\it{Upper Right:}} Doppler map obtained from difference imaging at H$\alpha$-core $\pm700$~m{\AA} wavelengths. The colour table 
indicates derived velocities in kms$^{-1}$, while the black circles mark the expected location of the magnetic-field footpoints 
described in Figure~\ref{f1}. {\it{Bottom:}} Normalized intensity (H$\alpha$ wing: solid line; Ca-K: dashed line) 
plotted as a function of time. Each intensity value is an average derived 
from within the box outlined by the dashed black lines in the upper-right panel. Intensity fluctuations in the minutes 
prior to the H$\alpha$ microflare are consistent with observations of Ellerman bombs (indicated with red arrows). 
\label{f2}}
\end{figure*}

The maximum MBP separation distance of $\approx$4400~km is achieved at 08:38~UT. During the next 4~min, both MBPs remain 
stationary and exhibit no fluctuations in their observed size. However, during this time, line-of-sight 
velocity oscillations are created along a curved path linking the two MBPs (upper-left panel of Fig.~\ref{f2}). These 
photospheric velocities are derived from 36~s cadence Doppler shifts of the H$\alpha$ line bisector positioned at $\pm$700~m{\AA} 
from the core of a rest H$\alpha$ profile, and constitute four complete periods with amplitude $\pm$5~kms$^{-1}$ 
and wavelength $\approx$1000~km. However, due to the telescope pointing being directed away from the centre of the solar 
disk, the cos $\theta \approx 0.8$ angle must be taken into consideration when determining absolute velocities. 
Since MBPs in the photosphere often show negligible inclination angles to the solar surface 
\citep[]{San94}, 
incorporation of the viewing angle provides an absolute velocity 
amplitude of $\pm$6~kms$^{-1}$ normal to the solar surface. 
Furthermore, simultaneous Doppler velocities are also established using the methods detailed in \cite{Sue95}, 
providing a Doppler map, $Dopp$, given by $Dopp = (C_{b} - C_{r}) / (C_{b} + C_{r} + 2)$,
where $C_{b}$ and $C_{r}$ are contrast images obtained using the relation $C = (I-I_{a})/I_{a}$. The $I_{a}$ 
corresponds to the average intensity over the entire dataset, while the subscripts $b$ and $r$ correspond to 
the wavelengths at H$\alpha$-core $-700$~m{\AA} and H$\alpha$-core $+700$~m{\AA}, respectively. These Doppler 
maps also reveal an identical velocity oscillation characterized by an amplitude of $\pm$5~kms$^{-1}$ 
($\pm$6~kms$^{-1}$ with viewing-angle correction) and a 
wavelength $\approx$1000~km (upper-right panel of Fig.~\ref{f2}). It is interesting to note that the 
velocity oscillation crosses the path of the magnetic neutral line derived from full disk {\sc{mdi}} 
magnetograms. Theoretical modelling by \cite{Vek05} 
has shown how quasi-static perturbations to an initially stable 
magnetic-field configuration may induce magnetic reconnection at the system's neutral X-point. Indeed, \cite{Vek98} 
have shown that even a weak external deformation of the magnetic-field lines can produce a 
substantial magnetic relaxation. As a result, flare sites are often observed to 
coincide with the magnetic neutral line \citep[]{Din03, Jes08}.

What exactly causes a deformation of the magnetic-field lines is open to interpretation. It has 
been suggested that Ellerman bombs \citep[]{Ell17}, 
which occur at the footpoints of magnetic loops \citep[]{Den95}, 
may trigger 
bi-directional velocity disturbances in regions of weak fluxes of mixed polarity \citep[]{Cha98}. 
Observations of 
such phenomena have recently been presented by \cite{Mad09}, 
where the reported observational signatures of Ellerman bombs 
include symmetric intensity enhancements (rise and decay times are approximately equal), 
and numerous up- and down-flows detected in the wings of the H$\alpha$ profile. 
Typical sizes of Ellerman bombs are $\approx$$1''$--$2''$ in diameter \citep[]{Nin98} 
and they evolve with lifetimes in the range 4--20~min \citep[]{Den95}. 
Furthermore, \cite{Qiu00} 
suggest that these events 
evolve with a phase difference between photospheric and chromospheric observations as a result of upward-moving plasma, 
triggered by the Ellerman bomb at a height $<$350~km.

From the lower panel of Figure~\ref{f2}, it is clear that repeated H$\alpha$-wing intensity enhancements 
are visible in the lead up to the H$\alpha$ microflare event. Using an integration area defined by the dashed black lines in the 
upper-right panel of Figure~\ref{f2}, these intensity fluctuations provide a symmetric profile enhancement of $\approx$12\% 
to the background quiescent flux. This is consistent with previous measurements related to Ellerman bombs, where 
\cite{Geo02} 
detect intensity enhancements of $\approx$5--30\%. However, if the integration areas are modified so that 
only the flux of the MBPs are considered, the intensity enhancements increase dramatically to $\approx$43\%, 
higher than reported in \cite{Zac87}. 
This larger intensity contrast may be a result of the high 
spatial resolution of our data, which allows fine-scale structures to be resolved without the 
intensity-smearing associated with less-favorable spatial resolution. Furthermore, the physical size of the MBPs 
are $\approx$1$''$ in diameter, remaining consistent with the dimensions of typical Ellerman bomb 
sites \citep[]{Nin98}. 
The MBPs associated with Ellerman bomb locations are often observed to exhibit proper motions parallel to the surface 
of the Sun \citep[]{Den97}. 
These motions can frequently exceed 1~kms$^{-1}$ in regions surrounding sunspots \citep[]{Geo02}. 
The lower panel of Figure~\ref{f1} demonstrates proper motion of 
1--1.8~kms$^{-1}$ for the MBPs under investigation. 
Thus, we deem these symmetric intensity fluctuations visible in the lower panel of Figure~\ref{f2} to 
be the signature of Ellerman bombs, which generate the periodic velocity perturbations observed along the loop 
structure. In addition, it has been suggested that the production of such bi-directional velocity flows may act as a 
precursor to the presence of magnetic reconnection \citep[]{Der94}. 
It should be noted that the 
periodic velocity perturbations evaluated here correspond to significant blue and red shifts ($\pm$6~kms$^{-1}$) of 
the H$\alpha$ line profile bisectors, yet remain below the sound speed in the photosphere.

An important parameter for forced reconnection is the magnitude of the driving displacement, $\delta$. 
This is the maximum distance by which the boundary is perturbed. Forced reconnection can take place in 
various configurations, but assuming the simplest scenario of reconnection occurring along the vertical field 
above a neutral line, the relevant displacement is horizontal. 
In this example, the average 
horizontal velocity is estimated to be $\approx$7~kms$^{-1}$, which, over a 4 minute time interval, gives a displacement of 
about 1600~km. Comparing this with the separation distance of 4400~km gives a normalised displacement 
$\delta/a \approx 1/3$. As a result, forced reconnection is expected to be strong and nonlinear.

At 08:42~UT, the periodic velocity perturbation linking the two MBPs gives way to a substantial H$\alpha$ microflare event, 
visible in both the H$\alpha$-wing and Ca-K data sets. The flare ribbon directly links the two MBPs, and closely outlines 
the location of the velocity oscillation observed immediately prior to the H$\alpha$ microflare (lower-right panel 
of Fig.~\ref{f1}). 
Typically, reconnection sites originate 
in locations of strong magnetic field gradient, where the field strength changes abruptly over a very short distance. 
Thus, we believe that where a significant intensity enhancement of 63\% above the 
quiescent mean (peak intensity from lower panel in Fig.~\ref{f2}) is found and a co-spatial magnetic neutral line is 
observed, is the location where forced magnetic reconnection occurs. An intensity enhancement of 63\% is consistent with 
previous microflare observations and theoretical predictions \citep[]{Hud91, Jes07}.

As predicted by \cite{Qiu00}, 
we utilize the high-temporal resolution of our data set (H$\alpha$ = 9~s; Ca-K = 2.5~s) 
to determine a clear phase shift 
in the light curves generated for photospheric (H$\alpha$ wing) and lower-chromospheric (Ca-K) observations. A 
time delay of $\approx$3~min is visible in the lower panel of Figure~\ref{f2}, and applies to both the signatures of 
Ellerman bombs (visible at times of 6.5 and 10~min in the H$\alpha$-wing light curve) and the H$\alpha$ microflare event (at a time 
of 14~min in the H$\alpha$-wing light curve). The H$\alpha$ line-bisector velocities determined at the time of peak 
H$\alpha$ microflare flux demonstrate values of $\approx$6~kms$^{-1}$. 
In the lower solar atmosphere where radiation is the dominant 
energy loss mechanism for flares \citep[]{Ems81}, 
magnetic reconnection often leads to isotropic heating 
of the plasma \citep[]{Mac89}. 
Thus, we can use the H$\alpha$ line-bisector velocities of $\approx$6~kms$^{-1}$ 
as a good estimate of the flare-induced plasma velocity normal to the solar surface.
As a result, for plasma travelling upwards towards the layer of the solar atmosphere seen in Ca-K wavelengths 
\citep[1200~km;][]{Bee69}, 
a time delay of $\approx$3~min and a velocity of $\approx$6~kms$^{-1}$ provides a traversed 
distance of $\approx$1000~km. Subtracting this value from the formation height of Ca-K provides a reconnection altitude of 
$\sim$200~km above the solar surface, consistent with the formation height of the H$\alpha$ wing ($<$300~km). 
In this regime, upward flowing, hot plasma, generated at the 
reconnection altitude of $\sim$200~km, is visible in H$\alpha$-wing observations ($<$300~km) immediately, before being 
detectable at Ca-K altitudes ($\approx$1200~km), after a time delay of $\approx$3~min.
This derived H$\alpha$ microflare altitude is consistent with lower-atmospheric reconnection 
models presented in \cite{Din99}.

It is interesting to note that the H$\alpha$ microflare intensity enhancements are not visible in the core of the H$\alpha$ line. 
A possible interpretation may be the magnetic-field lines connecting the two MBPs not reaching sufficiently 
high into the solar atmosphere to be seen in H$\alpha$-core observations. 
Another possibility, as noted by \cite{Che06}, 
is that the high densities found in the lower solar atmosphere, 
coupled with a partially 
ionized and highly collisional plasma, means that any energy released through magnetic reconnection is promptly 
consumed, preventing accelerated plasma from traversing large distances as they would in the corona. With an 
H$\alpha$-core formation height of $\approx$1500~km \citep[]{Ver81}, 
the lack of microflare signatures at this atmospheric 
height may be due to the hot plasma surrounding the reconnection site being rapidly cooled before it reaches these elevated 
altitudes. 

The evolution of the H$\alpha$ microflare continues over the next 9~min, during which the observed intensity 
gradually decays back to its quiescent value (see lower panel of Fig.~\ref{f2}). Three minutes after 
the initial impulsive phase of reconnection, an abundance of red-shifted 
plasma, particularly in locations surrounding the MBPs, is detected with velocities exceeding 6~kms$^{-1}$ 
(7.5~kms$^{-1}$ normal to the solar surface with viewing-angle correction). These 
downward velocity trends are consistent with \cite{Din95}, 
who describe how downward mass motions appear near 
the extreme edges of flare ribbons. 

Since, as discussed above, the
forced reconnection is strongly driven, this may result in the nonlinear phase predominating. The characteristic 
time-scale, $\tau_{N}$, of nonlinear forced reconnection is, $\tau_{N} \equiv (\tau_{0} \tau_{R} \tau_{A}^{3})^{1/5}$ \citep[]{Wan96}, 
where $\tau_{0} = a/v_0$ is the driving time-scale, $\tau_{R} = a^2/\eta$ is the resistive time-scale 
with $\eta$, the magnetic diffusivity, defined as $\eta = 1/\mu_{0} \sigma$, and 
$\tau_{A}$ is the Alfv{\'{e}}n time-scale ($a$ is the characteristic length-scale of the initial current sheet 
and $v_0$ the driving velocity). Adopting values corresponding to the above observations, i.e. 
$a \approx 4400$~km, $T \approx 5000$~K, $n = 2 \times 10^{21}$~m$^{-3}$, $B = 500$~G and 
$\sigma = 7.4 \times 10^{-4}~T^{3/2}$,  
we establish $\tau_{N} \approx 87$ minutes, assuming purely classical resistivity \citep[][]{Spi56}. 
This is somewhat longer than the observed reconnection time-scale of about 
9~minutes, although of the correct order-of-magnitude.  However, the derived time-scale is strongly dependent on the assumed 
current sheet length-scale, $a$. If $a$ is taken to be the gravitational scale-height \citep[following][]{Lit99b}, 
which is $\sim$100~km, a much more rapid reconnection time-scale is obtained. Furthermore, 
any effects due to anomalous resistivity will also significantly reduce the  
reconnection time scale. The energy release associated with this event will 
scale as the total magnetic energy multiplied by $(\frac{\delta}{a})^{2}$ \citep[]{Vek98}, 
where $\delta$, as mentioned above, is the magnitude of the boundary disturbance. 
Based on a loop of length 4000~km and diameter 725~km, this gives an order-of-magnitude energy release
of $10^{22}$~J, which is consistent with previous studies of microflare energetics \citep[e.g.][]{Han08}.

This event has been classified as an H$\alpha$ microflare, due to a lack of detectable X-ray emission. 
However, the impulsive nature of the H$\alpha$ lightcurve, accompanied by its derived energy release, 
duration, occurrence at a magnetic neutral line, associated plasma acceleration, and correlation with solar activity, 
suggests that this phenomena may have more in common with typical X-ray microflares than first anticipated. A 
lack of detectable X-ray emission may simply be due to the reconnection process occurring deep within the lower 
solar atmosphere, where the associated energy release is promptly consumed, producing either no X-ray emission, 
or emission so faint that current satellites cannot detect it.

\section{Concluding Remarks}
\label{conc}

The unprecedented spatial and temporal resolution data acquired with the Swedish Solar Telescope has revealed a small-scale 
impulsive brightening in H$\alpha$ and Ca-K, which we interpret as microflare activity driven by forced reconnection. A 
progressive separation of two MBPs, followed by a periodic velocity perturbation, 
results in magnetic reconnection at the location of a polarity inversion line. Multiple Ellerman bombs are 
observed in the vicinity of the MBPs using both H$\alpha$ and Ca-K imaging, which are deemed to trigger 
periodic velocity oscillations along the loop structure connecting the MBPs. 
Bi-directional velocity perturbations 
created by the Ellerman bombs reveal amplitudes of $\pm$6~kms$^{-1}$ and a wavelength $\approx$1000~km. 
We interpret the generated velocity oscillations as the magnetic-field deformation necessary to trigger forced 
reconnection, thus demonstrating how small-scale reconnection (e.g. from an Ellerman bomb) can trigger a larger reconnection event. 
Furthermore, the resulting magnetic reconnection occurs co-spatially with the magnetic neutral line at an 
estimated atmospheric height of $\sim$200~km.
Calculations suggest that the observed trends are best represented by forced 
reconnection models in a nonlinear regime, with an expected energy release of the order $10^{22}$~J.

\acknowledgments
DBJ thanks STFC for a Post-Doctoral Fellowship. 
FPK thanks AWE Aldermaston for the William Penney Fellowship. The SST is operated 
by the Institute for Solar Physics of the Royal Swedish Academy of Sciences. These observations have 
been funded by OPTICON, within the Research Infrastructures of FP6. This work is 
supported by EOARD.


\begin{thebibliography}{}
\bibitem[Beebe \& Johnson(1969)]{Bee69}
Beebe, H.~A., \& Johnson, H.~R., 1969, \solphys, 10, 79
\bibitem[Brown(1971)]{Bro71}
Brown, J.~C., 1971, \solphys, 18, 489
\bibitem[Bulanov \& Sakai(1998)]{Bul98}
Bulanov, S.~V., \& Sakai, J.-I., 1998, \apjs, 117, 599
\bibitem[Chae et~al.(1998)]{Cha98}
Chae, J., Wang, H., Lee, C.-Y., Goode, P.~R., \& Schuehle, U., 1998, \apjl, 497, L109
\bibitem[Chen et~al.(2001)]{Che01}
Chen, P.~F., Fang, C., \& Ding, M.~D., 2001, {\it{Chinese Journal of Astronomy and Astrophysics}}, 1, 176
\bibitem[Chen \& Ding(2006)]{Che06}
Chen, Q.~R., \& Ding, M.~D., 2006, \apj, 641, 1217
\bibitem[Christe et~al.(2008)]{Chr08}
Christe, S., Hannah, I.~G., Krucker, S., McTiernan, J., \& Lin, R.~P., 2008, \apj, 677, 1385
\bibitem[Crockett et~al.(2009)]{Cro09}
Crockett, P.~J., Jess, D.~B., Mathioudakis, M., \& Keenan, F.~P., 2009, \mnras, 397, 1852
\bibitem[Denker et~al.(1995)]{Den95}
Denker, C., de Boer, C.~R., Volkmer, R., \& Kneer, F., 1995, \aap, 296, 567
\bibitem[Denker(1997)]{Den97}
Denker, C., 1997, \aap, 323, 599 
\bibitem[Dere(1994)]{Der94}
Dere, K.~P., 1994, {\it{Advances in Space Research}}, 14, 13 
\bibitem[Ding et~al.(1995)]{Din95}
Ding, M.~D., Fang, C., \& Huang, Y.~R., 1995, \solphys, 158, 81
\bibitem[Ding et~al.(1999)]{Din99}
Ding, M.~D., Fang, C., \& Yun, H.~S., 1999, \apj, 512, 454
\bibitem[Ding et~al.(2003)]{Din03}
Ding, M.~D., Chen, Q.~R., Li, J.~P., \& Chen, P.~F., 2003, \apj, 598, 683
\bibitem[Ellerman(1917)]{Ell17}
Ellerman, F., 1917, \apj, 46, 298
\bibitem[Emslie et~al.(1981)]{Ems81}
Emslie, A.~G., Brown, J.~C., \& Machado, M.~E., 1981, \apj, 246, 337
\bibitem[Georgoulis et~al.(2002)]{Geo02}
Georgoulis, M.~K., Rust, D.~M., Bernasconi, P.~N., \& Schmieder, B., 2002, \apj, 575, 506
\bibitem[Hannah et~al.(2008)]{Han08}
Hannah, I.~G., Christe, S., Krucker, S., Hurford, G.~J., Hudson, H.~S., \& Lin, R.~P., 2008, \apj, 677, 704
\bibitem[Hudson(1991)]{Hud91}
Hudson, H.~S., 1991, \solphys, 133, 357 
\bibitem[Jain et~al.(2005)]{Jai05}
Jain, R., Browning, P., \& Kusano, K., 2005, {\it{Physics of Plasmas}}, 12, 012904
\bibitem[Jess et~al.(2007)]{Jes07}
Jess, D.~B., McAteer, R.~T.~J., Mathioudakis, M., Keenan, F.~P., Andic, A., \& Bloomfield, D.~S., 2007, \aap, 476, 971
\bibitem[Jess et~al.(2008)]{Jes08}
Jess, D.~B., Mathioudakis, M., Crockett, P.~J., \& Keenan, F.~P., 2008, \apjl, 688, L119
\bibitem[Krucker et~al.(2002)]{Kru02}
Krucker, S., Christe, S., Lin, R.~P., Hurford, G.~J., \& Schwartz, R.~A., 2002, \solphys, 210, 445
\bibitem[Litvinenko(1999)]{Lit99a}
Litvinenko, Y.~E., 1999, \apj, 515, 435
\bibitem[Litvinenko \& Martin(1999)]{Lit99b}
Litvinenko, Y.~E., \& Martin, S.~F., 1999, \solphys, 190, 45 
\bibitem[Litvinenko et~al.(2007)]{Lit07}
Litvinenko, Y.~E., Chae, J., \& Park, S.-Y., 2007, \apj, 662, 1302 
\bibitem[Ma et~al.(1996)]{Ma96}
Ma, Z.~W., Wang, X., \& Bhattacharjee, A., 1996, {\it{Physics of Plasmas}}, 3, 2427
\bibitem[Machado et~al.(1989)]{Mac89}
Machado, M.~E., Emslie, A.~G., \& Avrett, E.~H., 1989, \solphys, 124, 303
\bibitem[Madjarska et~al.(2009)]{Mad09}
Madjarska, M.~S., Doyle, J.~G., \& de Pontieu, B., 2009, \apj, 701, 253 
\bibitem[Nindos \& Zirin(1998)]{Nin98}
Nindos, A., \& Zirin, H., 1998, \solphys, 182, 381 
\bibitem[Priest(1986)]{Pri86}
Priest, E.~R., 1986, {\it{Advances in Space Research}}, {\bf{6}}, 73 
\bibitem[Qiu et~al.(2000)]{Qiu00}
Qiu, J., Ding, M.~D., Wang, H., Denker, C., \& Goode, P.~R., 2000, \apjl, 544, L157
\bibitem[Roumeliotis \& Moore(1993)]{Rou93}
Roumeliotis, G., \& Moore, R.~L., 1993, \apj, 416, 386
\bibitem[Sanchez Almeida \& Martinez Pillet(1994)]{San94}
Sanchez Almeida, J., \& Martinez Pillet, V., 1994, \apj, 424, 1014
\bibitem[Spitzer(1956)]{Spi56}
Spitzer, L., 1956, {\it{Physics of Fully Ionized Gases}}, Interscience Publishing
\bibitem[Suematsu et~al.(1995)]{Sue95}
Suematsu, Y., Wang, H., \& Zirin, H., 1995, \apj, 450, 411 
\bibitem[Vekstein \& Jain(1998)]{Vek98}
Vekstein, G.~E., \& Jain, R., 1998, {\it{Physics of Plasmas}}, 5, 1506
\bibitem[Vekstein \& Bian(2005)]{Vek05}
Vekstein, G., \& Bian, N., 2005, \apjl, 632, L151
\bibitem[Vernazza et~al.(1981)]{Ver81}
Vernazza, J.~E., Avrett, E.~H., \& Loeser, R., 1981, \apjs, 45, 635 
\bibitem[Wang et~al.(1996)]{Wan96}
Wang, X., Ma, Z.~W., \& Bhattacharjee, A., 1996, {\it{Physics of Plasmas}}, 3, 2129
\bibitem[Zachariadis et~al.(1987)]{Zac87}
Zachariadis, T.~G., Alissandrakis, C.~E., \& Banos, G., 1987, \solphys, 108, 227
\end{thebibliography}
\end{document}